\documentclass[twocolumn, amssymb, amsmath, aps, superscriptaddress, showpacs, footinbib,  prb]{revtex4}
\pdfoutput=1
\usepackage{graphicx}
\usepackage{dcolumn}

\newcommand{\eff}{\text{eff}}
\newcommand{\AFM}{\text{AFM}}

\newcolumntype{C}{>{$}c<{$}}

\begin{document}

\title{Extension of the spin-1/2 frustrated square lattice model: \\the case of layered vanadium phosphates}
\author{Alexander A. Tsirlin}
\email{altsirlin@gmail.com}
\affiliation{Max Planck Institute for Chemical Physics of Solids, N\"{o}thnitzer
Str. 40, 01187 Dresden, Germany}
\affiliation{Department of Chemistry, Moscow State University, 119992 Moscow, Russia}
\author{Helge Rosner}
\email{Helge.Rosner@cpfs.mpg.de}
\affiliation{Max Planck Institute for Chemical Physics of Solids, N\"{o}thnitzer
Str. 40, 01187 Dresden, Germany}

\begin{abstract}
We study the influence of the spin lattice distortion on the properties of frustrated magnetic systems and consider the applicability of the spin-1/2 frustrated square lattice model to materials lacking tetragonal symmetry. We focus on the case of layered vanadium phosphates AA$'$VO(PO$_4)_2$ (\mbox{AA$'$ = Pb$_2$,} SrZn, BaZn, and BaCd). To provide a proper microscopic description of these compounds, we use extensive band structure calculations for real materials and model structures and supplement this analysis with simulations of thermodynamic properties, thus facilitating a direct comparison with the experimental data. Due to the reduced symmetry, the realistic spin model of layered vanadium phosphates AA$'$VO(PO$_4)_2$ includes four inequivalent exchange couplings: $J_1$ and $J_1'$ between nearest-neighbors and $J_2$ and $J_2'$ between next-nearest-neighbors. The estimates of individual exchange couplings suggest different regimes, from $J_1'/J_1$ and $J_2'/J_2$ close to~1 in BaCdVO(PO$_4)_2$, a nearly regular frustrated square lattice, to $J_1'/J_1\simeq 0.7$ and $J_2'/J_2\simeq 0.4$ in SrZnVO(PO$_4)_2$, a frustrated square lattice with sizable distortion. The underlying structural differences are analyzed, and the key factors causing the distortion of the spin lattice in layered vanadium compounds are discussed. We propose possible routes for finding new frustrated square lattice materials among complex vanadium oxides. Full diagonalization simulations of thermodynamic properties indicate the similarity of the extended model to the regular one with averaged couplings. In case of moderate frustration and moderate distortion, valid for all the AA$'$VO(PO$_4)_2$ compounds reported so far, the distorted spin lattice can be considered as a regular square lattice with the couplings $(J_1+J_1')/2$ between nearest-neighbors and $(J_2+J_2')/2$ between next-nearest-neighbors.
\end{abstract}

\maketitle

\section{Introduction}
Frustrated spin systems represent one of the actively developing topics in solid state physics. The vast interest in magnetic frustration originates from a number of unusual phenomena (spin-liquid ground state,\cite{lee2008} the formation of supersolid phases in high magnetic fields,\cite{supersolid} etc.) suggested by theory. The theoretical predictions challenge the experiment that, however, requires proper frustrated materials. The search for the respective compounds has been a long story in inorganic chemistry,\cite{frustrated1994,frustrated2001} and the problem turned out to be quite complex. To meet the theoretical predictions, one has to find a material that reveals frustrated geometry of magnetic atoms and presents spin degrees of freedom only to avoid any foreign effects (e.g., orbital ordering) tending to lift the frustration. A number of frustrated spin models still lack the proper realizations, especially for the case of spin-1/2, where the strongest quantum effects and the most interesting phenomena are expected. For other models, few appropriate materials are known and extensively studied. For example, the mineral herbertsmithite ZnCu$_3$(OH)$_6$Cl$_2$ was recently proposed as a spin-1/2 kagom\'e material.\cite{kagome} However, the actual physics of this compound is still debated due to the Cu/Zn antisite disorder and the presence of non-magnetic sites within the kagom\'e layers.\cite{kagome-2008} Other natural kagom\'e materials -- minerals kapellasite and haydeeite -- are proposed. Their structures do not suffer from the disorder effects, but the pure kagom\'e physics is again modified due to the non-negligible interactions beyond nearest neighbors.\cite{oleg}

Clearly, the task of finding an ideal frustrated material is hardly solvable at all. Therefore, it is instructive to examine which deviations from the ideal spin model may be allowed and, to a certain extent, do not qualitatively modify the properties of this model. Considering structural distortions and the resulting spin lattice distortions is especially attractive, since lots of known materials have low symmetry in contrast to the high geometrical symmetry of the most theoretically studied frustrated spin models.

The major part of the frustrated magnetic materials are the so-called geometrically frustrated magnets. In these systems, the frustration arises due to the competition of equivalent exchange couplings: in the most simple case, antiferromagnetic (AFM) couplings on a triangle. Then, any structural distortion should inevitably reduce this competition hence reducing the frustration. In the other group of the frustrated materials, the competing interactions are inequivalent but their topology and magnitudes can be tuned so that the strong quantum fluctuations destroy the long-range ordering, similar to the geometrically frustrated magnets. The latter group looks more favorable to tolerate the structural distortions, since the modification of the spin lattice can probably be balanced by the ratios of the competing interactions hence preserving the strong frustration. To study this issue in more detail, we focus on a specific model -- spin-1/2 frustrated square lattice (FSL) -- and consider a number of recently proposed FSL materials.

\begin{figure}
\includegraphics[scale=1]{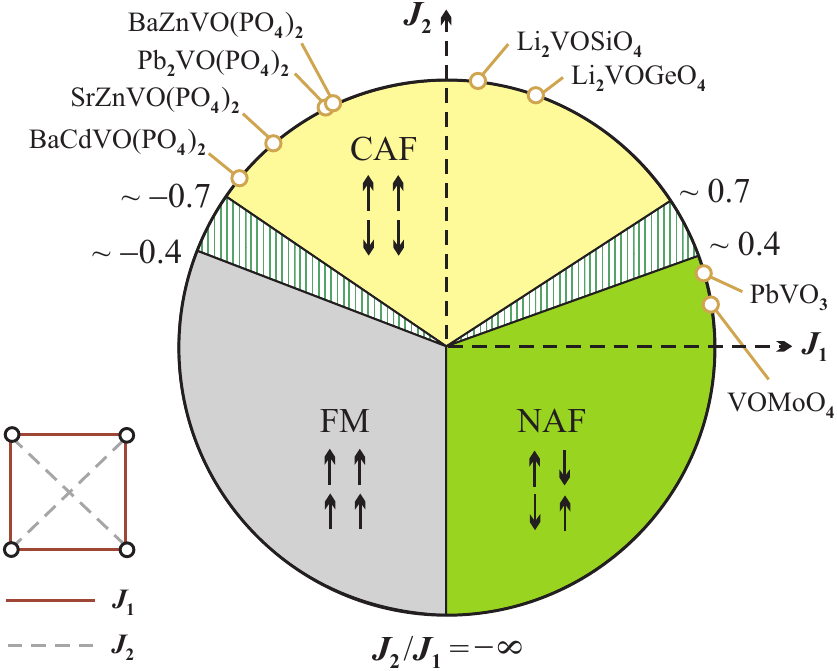}
\caption{\label{fig_diagram}(Color online)
Phase diagram of the FSL model\cite{shannon2004} and the respective model compounds (see text for references). The inset shows the regular FSL with the NN ($J_1$) and NNN ($J_2$) couplings denoted by solid and dashed lines, respectively.
}
\end{figure}

The ideal (regular) FSL model assumes two competing interactions: the nearest-neighbor (NN) interaction $J_1$ running along the side of the square and the next-nearest-neighbor (NNN) interaction $J_2$ running along the diagonal of the square (see the inset of Fig.~\ref{fig_diagram}). The phase diagram (Fig.~\ref{fig_diagram}) reveals three ordered phases (ferromagnet, N\'eel antiferromagnet, and columnar antiferromagnet) and two critical regions around $J_2/J_1\simeq\pm 0.5$, where a spin-liquid ground state is expected.\cite{frustrated-book,shannon2004} Recent theoretical studies also considered the extended model with inequivalent NN couplings $J_1$ and $J_1'$ (Refs.~\onlinecite{nersesyan2003,starykh2004,sindzingre2004,moukouri2006,bishop2008}). This spatial anisotropy tends to narrow the critical region\cite{foot1} and to destroy it completely at a certain value of $J_1'/J_1$. Therefore, the distortion is not favorable for the frustration, but the specific geometry of the lattice enables to preserve the strong frustration and the resulting critical region at moderate distortion. Below (Sec.~\ref{simulation}), we will show that the anisotropy of the NNN couplings on the square lattice should have an even weaker (and, likely, opposite) effect on the frustration.

Experimental studies of the FSL systems have utilized a number of model compounds. First reports focused on Li$_2$VOXO$_4$ (X = Si, Ge) materials that lay far away from the critical regions and did not show any specific properties caused by the frustration.\cite{melzi2000,melzi2001,helge2002,helge2003,misguich2003,bombardi2004} Later on, the vanadyl molybdate VOMoO$_4$ was found to reveal unusual structural changes upon cooling, and this effect was tentatively ascribed to the magnetic frustration despite the frustration ratio $J_2/J_1$ was quite small (likely, $J_2/J_1<0.2$).\cite{carretta2002,bombardi2005} The region of ferromagnetic (FM) $J_1$ -- AFM $J_2$ was accessed by studying the layered vanadium phosphate Pb$_2$VO(PO$_4)_2$ (Refs.~\onlinecite{kaul2004,enrique,skoulatos2007}) and the related AA$'$VO(PO$_4)_2$ compounds with AA$'$ = SrZn and BaZn.\cite{enrique,skoulatos2008} Quite recently, we proposed two more compounds, PbVO$_3$ (Refs.~\onlinecite{pbvo3} and \onlinecite{azuma2008}) and BaCdVO(PO$_4)_2$ (Ref.~\onlinecite{bacdvp2o9}), that lay very close to the critical regions at $J_2/J_1=0.5$ and $-0.5$, respectively. Interestingly, the latter material lacks the tetragonal symmetry and reveals a distorted FSL. Nevertheless, we succeeded to observe two effects predicted for the regular FSL: the suppression of the specific heat maximum\cite{shannon2004} and the pronounced bending of the magnetization curve.\cite{thalmeier2008} 

In other systems, the problem of the spin lattice distortion may also be crucial. The layered copper oxychloride (CuCl)LaNb$_2$O$_7$ was recently proposed as a promising FSL material, lacking long-range magnetic order.\cite{kageyama2005} However, careful studies indicated the structural distortion\cite{yoshida2007} that completely changed the underlying spin model and precluded from any interpretations within the FSL framework.\cite{cucl} Thus, it is important to understand whether the FSL model can be applied to BaCdVO(PO$_4)_2$ and, more generally, to the low-symmetry materials. Below, we study this issue in detail, discuss the whole family of the AA$'$VO(PO$_4)_2$ compounds (hereinafter, we imply that AA$'$ = Pb$_2$, SrZn, BaZn, and BaCd), and provide quantitative estimates for the spin lattice distortion. Our approach combines several computational methods (band structure calculations, subsequent analysis of the exchange couplings, and model simulations) in order to analyze magnetic interactions in these systems, derive the proper spin model, and facilitate the comparison with the experimental data. We show that the structural distortion in the AA$'$VO(PO$_4)_2$ compounds is a minor effect as compared to the frustration, and the FSL description holds.

The outline of the paper is as follows. We start with an analysis of the crystal structures in Sec.~\ref{structure} and review the computational methods employed in our work (Sec.~\ref{methods}). Then, we address several problems: (i) the realistic spin model and the magnitude of the distortion (Sec.~\ref{distortion}); (ii) structural factors that influence the distortion of the spin lattice (Sec.~\ref{factors}); and (iii) thermodynamic properties of the extended model (Sec.~\ref{simulation}). In Sec.~\ref{discussion}, we discuss the results of our study that suggests an accurate (and, within the Heisenberg model, exact) way to treat the FSL-like spin systems of the AA$'$VO(PO$_4)_2$ phosphates. We also present a more general recipe for finding strongly frustrated square lattices in the compounds with similar topology of the magnetic layer, being generic for most of the FSL materials. Finally, we show how the distortion of the square lattice affects thermodynamic properties of the model and the magnitude of the frustration. 

\begin{figure*}
\includegraphics[scale=1]{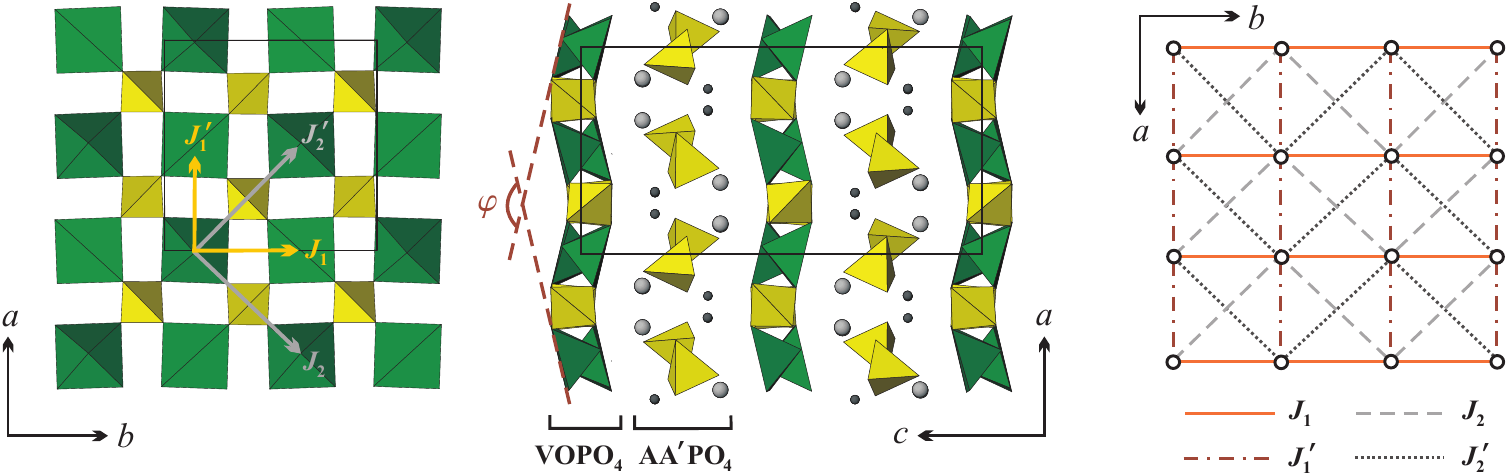}
\caption{\label{fig_layer}(Color online)
Crystal structure of the AA$'$VO(PO$_4)_2$ compounds and the underlying spin model. The left panel shows a single [VOPO$_4$] layer. The middle panel presents the stacking of the [VOPO$_4$] layers and the [AA$'$PO$_4$] blocks as well as the angle $\varphi$ measuring the layer buckling: larger and smaller spheres denote the A and A$'$ cations, respectively. The right panel shows the magnetic interactions in the [VOPO$_4$] layers (compare with the regular model in the inset of Fig.~\ref{fig_diagram}): solid, dash-dotted, dashed, and dotted lines indicate $J_1,J_1',J_2$, and $J_2'$, respectively.
}
\end{figure*}

\section{Crystal structures and experimental results}
\label{structure}
Most of the vanadium-based FSL compounds reported so far reveal similar [VOXO$_4$] magnetic layers. These layers are built of VO$_5$ pyramids and XO$_4$ tetrahedra (see left panel of Fig.~\ref{fig_layer}) with X being a non-magnetic cation (P, Si, Ge, or Mo$^{+6}$). The connections via the tetrahedra provide superexchange pathways for both NN and NNN couplings. Vanadium atoms have the steady oxidation state of +4 (implying the electronic configuration $3d^1$ and spin-1/2), while the valence of the X cation can be changed and controls the filling of the interlayer space. Thus, the interlayer space is empty in VOMoO$_4$ (Ref.~\onlinecite{vomoo4}), filled by lithium atoms in Li$_2$VOXO$_4$ (X = Si, Ge)\cite{millet1999} and filled by complex interlayer [AA$'$PO$_4$] blocks in the AA$'$VO(PO$_4)_2$ phosphates.\cite{srznvp2o9,baznvp2o9,shpanchenko} The magnetic layer is compatible with the tetragonal symmetry, hence the FSL model can be realized. Yet the overall symmetry of the structure is sometimes reduced due to the complex configuration of the interlayer block, and this is the case for the AA$'$VO(PO$_4)_2$ compounds.

The crystal structure of the AA$'$VO(PO$_4)_2$ phosphates (Fig.~\ref{fig_layer}) is fairly flexible with respect to the accommodation of different divalent metal cations A and A$'$ in the interlayer block. Most of the compounds combine a larger (Ba or Sr) and a smaller (Cd or Zn) cation, yielding the orthorhombic symmetry of the structure (space group $Pbca$).\cite{srznvp2o9,baznvp2o9} Yet it is also possible to use the same cation (Pb) for A and A$'$, and the resulting structure has monoclinic symmetry (space group $P2_1/c$).\cite{shpanchenko} Different symmetries produce slightly different geometries of the magnetic [VOPO$_4$] layers, hence different types of the spin lattice distortion should be expected. In the following, we will focus on the magnetic layer typical for the orthorhombic AA$'$VO(PO$_4)_2$ compounds with different A and A$'$ (Fig.~\ref{fig_layer}). To ease the comparison to monoclinic Pb$_2$VO(PO$_4)_2$, one should convert the crystal structure\cite{shpanchenko} to the non-standard setting $P2_1/b$ with $a$ being the monoclinic axis (in this setting, the layer corresponds to one shown in Fig.~\ref{fig_layer}). Basically, the whole discussion of the spin lattice distortion is applicable to Pb$_2$VO(PO$_4)_2$, but due to the lower lattice symmetry this compound has one additional feature: two inequivalent NN couplings along the $b$ axis. However, our estimates show that these couplings nearly match. Therefore, the spin system of Pb$_2$VO(PO$_4)_2$ can be adequately described with four parameters similar to the other AA$'$VO(PO$_4)_2$ compounds.

\begin{table}[!b]
\renewcommand{\arraystretch}{1.5}
\setlength{\belowcaptionskip}{0.2cm}
\caption{\label{tab_experiment}
Experimental exchange couplings ($J_1^{\exp},J_2^{\exp}$) evaluated within the regular FSL model and the resulting frustration ratios in the AA$'$VO(PO$_4)_2$ compounds
}
\begin{ruledtabular}
\begin{tabular}{ccccc}
  AA$'$ & $J_1^{\exp}$ (K) & $J_2^{\exp}$ (K) & $J_2^{\exp}/J_1^{\exp}$ & Ref. \\
  BaCd & $-3.6$ & $3.2$ & $-0.9$ & \onlinecite{bacdvp2o9} \\
  Pb$_2$ & $-5.1$ & $9.4$ & $-1.8$ & \onlinecite{enrique,kaul2004} \\
  BaZn & $-5.0$ & $9.3$ & $-1.9$ & \onlinecite{enrique} \\
  SrZn & $-8.3$ & $8.9$ & $-1.1$ & \onlinecite{enrique} \\
\end{tabular}
\end{ruledtabular}
\end{table}

The lack of the tetragonal symmetry in the AA$'$VO(PO$_4)_2$ phosphates gives rise to four different interactions in the magnetic [VOPO$_4$] layers. According to Fig.~\ref{fig_layer}, we label the NN interactions as $J_1$, $J_1'$ and the NNN interactions as $J_2, J_2'$. Previous experimental works implicitly assumed the ideal FSL model with $J_1=J_1'$ and $J_2=J_2'$ as a natural, albeit therein unjustified, approximation.\cite{enrique,kaul2004,bacdvp2o9} In the previous studies, experimental data on the magnetic susceptibility and the specific heat were fitted with high-temperature series expansions (HTSE) for the regular FSL\cite{helge2003} to yield the effective couplings $J_1^{\exp}$ and $J_2^{\exp}$ (Table~\ref{tab_experiment}). The FM $J_1$ -- AFM $J_2$ regime of the FSL was further supported by analyzing field dependence of the magnetization of BaCdVO(PO$_4)_2$ (Ref.~\onlinecite{bacdvp2o9}) and the ground states of Pb$_2$VO(PO$_4)_2$ and SrZnVO(PO$_4)_2$ (Refs.~\onlinecite{skoulatos2007} and~\onlinecite{skoulatos2008}). 

Within the phenomenological approach, the consistent interpretation of the experimental results justifies a posteriori the application of the ideal FSL model to the AA$'$VO(PO$_4)_2$ compounds. Of course, a microscopic approach requires a more careful consideration of all the four inequivalent exchange couplings in the magnetic [VOPO$_4$] layers. However, it is quite difficult (at least, experimentally) to go beyond the regular FSL description due to the lack of theoretical results for the extended $J_1-J_1'-J_2-J_2'$ model. In the following, we address the problem using computational methods. These methods are known to provide a reliable microscopic description of complex spin systems, including the FSL compounds Li$_2$VOXO$_4$ (Refs.~\onlinecite{helge2002} and \onlinecite{helge2003}). Yet we also refer to the phenomenological results and show that at sufficiently high temperatures the distorted FSL can be considered as a regular FSL with effective, averaged NN and NNN couplings.

\section{Methods and modeling}
\label{methods}
Scalar-relativistic band structure calculations were performed within the full-potential local-orbital scheme (FPLO7.00-27),\cite{fplo} and the exchange-correlation potential by Perdew and Wang\cite{perdew} was applied. Our calculations employed experimental crystal structures and a number of modified and model structures, as described below. Different $k$-meshes were used depending on the size and the geometry the unit cell. In all the calculations, the convergence with respect to the $k$-mesh was carefully checked.

To evaluate the exchange couplings in the AA$'$VO(PO$_4)_2$ compounds, we use two different approaches. First, local density approximation (LDA) calculations are performed. These calculations enable to select relevant states and to estimate hopping parameters ($t$) for the respective bands by fitting these bands with a tight-binding (TB) model. The LDA calculations fail to reproduce the strong correlation effects in the vanadium $3d$ shell; therefore, the correlations are included on a model level. The hoppings are introduced to an extended Hubbard model with the effective on-site Coulomb repulsion $U_{\eff}=4.5$ eV (this value is representative for vanadium oxides, see Refs.~\onlinecite{helge2002}, \onlinecite{helge2003}, \onlinecite{ag2vop2o7}, and \onlinecite{sr2v3o9}). The strongly correlated limit $t\ll U_{\eff}$ and the half-filling regime justify the reduction to the Heisenberg model for the low-lying excitations. Then, AFM contributions to the exchange couplings are estimated as $J_i^{\AFM}=4t_i^2/U_{\eff}$. Within this approach, all the possible AFM couplings are evaluated. 

Second, we consider the correlation effects within the self-consistent calculations and employ the local spin density approximation (LSDA)+$U$ method. Total energies for a number of ordered spin configurations are mapped onto the classical Heisenberg model to yield the estimates for both FM and AFM couplings, hence supplementing the TB analysis. LSDA+$U$ treats correlation effects in a mean-field approximation and uses two input parameters, $U_d$ and $J_d$, to describe the on-site Coulomb repulsion and the intraatomic exchange, respectively. Since the exchange parameter has minor influence on the results, we fix $J_d=1$ eV as a representative value. Yet the choice of the repulsion parameter $U_d$ may be crucial, especially in case the exchange couplings are weak (see Ref.~\onlinecite{cavpo} for an instructive example). To reduce the ambiguity related to the choice of $U_d$, we thoroughly compare the computational results with the experimental data, as further discussed in Sec.~\ref{distortion}. 

The LDA calculations employed the full symmetry of the crystal structures, 112-atom orthorhombic unit cells, and the $k$-mesh of 256 $k$-points with 75 points in the irreducible part of the first Brillouine zone (IBZ). In case of Pb$_2$VO(PO$_4)_2$, the 56-atom monoclinic unit cell and a mesh of 512 $k$-points (170 in IBZ) were used. 

To realize different spin orderings within the LSDA+$U$ calculations, one has to reduce the crystal symmetry and, in some cases, to extend the unit cell. The evaluation of the four exchange couplings in the AA$'$VO(PO$_4)_2$ compounds requires the doubling of the unit cell in the $a$ direction, hence 224-atom unit cells should be constructed. For such unit cells, full-potential calculations are extremely time-consuming and, likely, not accurate enough for a reliable evaluation of the rather small exchange constants.\cite{foot7} Therefore, in our LSDA+$U$ calculations we simplify the crystal structures of the AA$'$VO(PO$_4)_2$ compounds and construct a number of modified structures. The idea resembles our study of Ag$_2$VOP$_2$O$_7$ (Ref.~\onlinecite{ag2vop2o7}): the leading magnetic interactions take place in the \mbox{V--P--O} layers, hence it is essential to use the correct geometry of the layer, while the stacking of the layers and the filling of the interlayer space have minor effect on the leading exchange couplings. 

To build the simplified structures of the AA$'$VO(PO$_4)_2$ compounds, we keep the exact geometry of the [VOPO$_4$] layers, stack these layers one onto another, and fill the interlayer space with lithium atoms, providing the proper charge balance. The resulting composition is LiVOPO$_4$. The interlayer separation is fixed at 6.5 \r A to achieve realistic, sufficiently weak interlayer hoppings (below 2~meV). To justify the structure simplification, we perform the TB analysis. The difference between the respective hoppings in the experimental and modified structures does not exceed 5~\%, implying an error below 10~\% for the exchange couplings. Such an error is definitely acceptable for the further LSDA+$U$ calculations. Basically, the structure simplification provides a promising computational approach to the magnetic properties of low-dimensional spin systems with complex crystal structures. In our LSDA+$U$ calculations, we use 64-atom supercells [$2a\times b\times c\,(=6.5$ \r A)] with triclinic symmetry (space group $P1$) and a mesh of 108 $k$-points.

Finally, we also performed a number of LDA calculations for model structures in order to study the influence of individual structural changes on the spin lattice distortion. The model structures were built similar to the simplified structures described above. The initial geometry was taken from the structure of $\alpha_{\text{I}}$-LiVOPO$_4$ that includes regular [VOPO$_4$] layers separated by Li cations.\cite{livopo4} Then, we introduced a number of structural distortions and checked the changes of the exchange couplings (see Sec.~\ref{factors} for details). The initial structure reveals a rather low interlayer separation of 4.45 \r A and yields sizable interlayer interactions. To reduce the interlayer interactions and to properly emulate the two-dimensional (2D) character of the FSL compounds, we increased the interlayer spacing up to 6.5 \r A. We also reduced the symmetry down to the orthorhombic space group $Pbma$ ($Pbcm$ in the standard setting) that allowed the distortions of the [VOPO$_4$] layer. A $k$-mesh of 4096 points (729 in the IBZ) was used. 

Full diagonalization (FD) simulations were performed for the $N=16\ (4\times 4)$ cluster using the ALPS simulation package.\cite{alps2} Basic thermodynamic quantities (magnetic susceptibility and specific heat) were evaluated by an internal procedure of the program. In general, the FD simulations suffer from finite-size effects, because current computational facilities do not allow to perform the calculations for large clusters. The presently available cluster size (normally, 16 or 20 sites) is sufficient to obtain the quantitatively correct information on thermodynamic properties of one-dimensional spin systems only. The accuracy of the FD simulations for two-dimensional systems is challenged by the experimental data and the results of other simulation techniques.\cite{enrique,bacdvp2o9,foot6} Yet, the FD simulations are able to provide qualitatively correct trends upon the change of the model parameters (e.g., the change of the frustration ratio in the FSL, see Ref.~\onlinecite{shannon2004}). Keeping in mind these considerations, we restrict ourselves to the analysis of the relative changes in the thermodynamic properties, as the spin lattice is distorted. Investigation of the ground state and quantitative simulation of thermodynamic properties for the extended FSL model are clearly beyond the scope of the present paper that intends to stimulate further studies of the problem.

\section{Exchange couplings in AA$'$VO(PO$_4)_2$}
\label{distortion}

\subsection{LDA and tight-binding analysis}
\begin{figure}
\includegraphics[scale=1.0]{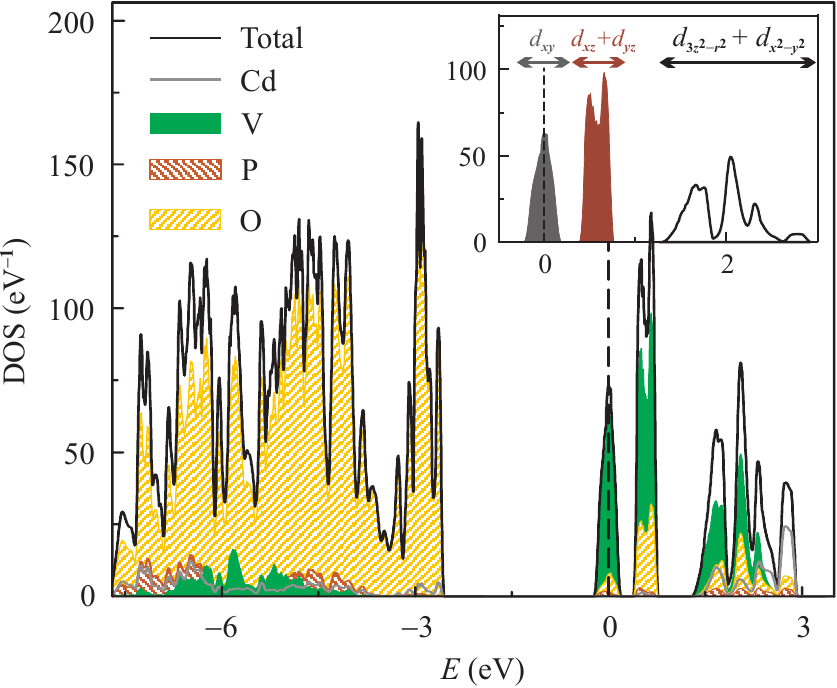}
\caption{\label{fig_dos}(Color online)
LDA density of states for BaCdVO(PO$_4)_2$ (the contribution of barium is not shown, because it is negligible in the whole energy range). The Fermi level is at zero energy. The inset shows the orbital resolved DOS for vanadium.
}
\end{figure}

In this section, we consider the band structures of the AA$'$VO(PO$_4)_2$ compounds and evaluate individual exchange couplings. The LDA density of states (DOS) for BaCdVO(PO$_4)_2$ is shown in Fig.~\ref{fig_dos}. This plot is representative for the whole family of the AA$'$VO(PO$_4)_2$ materials. The electronic structure resembles that of other vanadium phosphates.\cite{cavpo} The states below $-2.5$ eV are mainly formed by oxygen orbitals, while the states close to the Fermi level reveal predominant vanadium contribution. The vanadium bands are rather narrow and show $3d$-related crystal field levels (see the inset of Fig.~\ref{fig_dos}) as expected for the square-pyramidal or distorted octahedral coordination of V$^{+4}$ (Ref.~\onlinecite{ballhausen}). The lowest-lying vanadium states formed by $d_{xy}$ orbitals lie at the Fermi level. The respective orbitals are located in the basal planes of the square pyramids hence overlapping with $p$ orbitals of the basal oxygen atoms and facilitating exchange couplings in the $ab$ plane. The LDA energy spectrum is metallic in contradiction to the experimental green or yellow-green color that indicates insulating behavior. The lack of the energy gap is a typical failure of LDA due to the underestimate of strong electron correlations in the $3d$ shell. Applying LSDA+$U$, we find insulating spectra with an energy gap of $1.5-2.8$ eV depending on the $U_d$ value. The upper estimate is in reasonable agreement with the experimental sample colors.

\begin{figure*}
\includegraphics{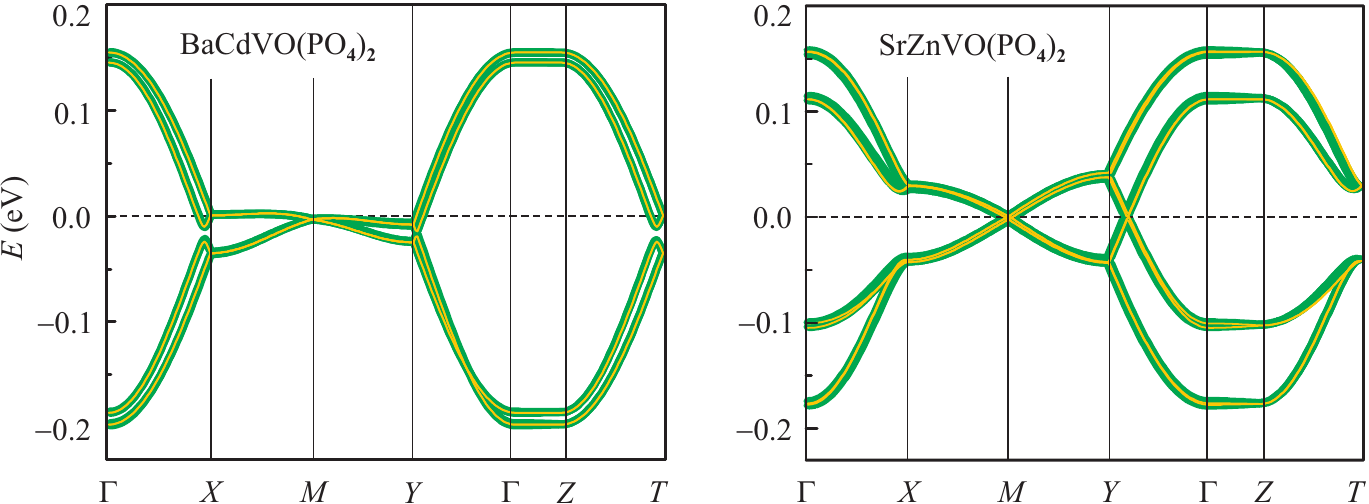}
\caption{\label{fig_bands}
(Color online) LDA band structure (thin light lines) and the fit of the tight-binding model (thick green lines) for BaCdVO(PO$_4)_2$ (left panel) and SrZnVO(PO$_4)_2$ (right panel). The Fermi level is at zero energy. The notation of the $k$ points is as follows: $\Gamma(0,0,0)$, $X(0.5,0,0)$, $M(0.5,0.5,0)$, $Y(0,0.5,0)$, $Z(0,0,0.5)$, and $T(0.5,0,0.5)$ (the coordinates are given along $k_x$, $k_y$, and $k_z$ in units of the respective reciprocal lattice parameters).
}
\end{figure*}

For the TB analysis, we select the eight half-filled $d_{xy}$ bands lying at the Fermi level (Fig.~\ref{fig_bands}). These bands originate from eight vanadium atoms in the unit cell of AA$'$VO(PO$_4)_2$. The unit cell includes two [VOPO$_4$] layers (see middle panel of Fig.~\ref{fig_layer}) and four vanadium atoms from each of the two layers. The layers are well separated by the [AA$'$PO$_4$] block; therefore, interlayer interactions are very weak, and the bands are close to double degeneracy in the whole Brillouin zone. The leading interlayer hopping is about 1 meV, implying the AFM interactions of about 0.01~K, well below the in-layer interactions (see Tables~\ref{tab_experiment} and~\ref{tab_hoppings}). In Fig.~\ref{fig_bands}, we show the band structures of BaCdVO(PO$_4)_2$ and SrZnVO(PO$_4)_2$ at the Fermi level. The two plots are rather similar in the overall behavior and in the energy scale, although notable differences are found near the $X$, $Y$, and $T$ points. The TB analysis suggests that the differences are mainly related to the NNN couplings $t_2$ and $t_2'$. 

\begin{table}[!hb]
\renewcommand{\arraystretch}{1.5}
\setlength{\belowcaptionskip}{0.2cm}
\caption{\label{tab_hoppings}
The hopping parameters $t_i$ (in meV) and the resulting magnitude of the distortion ($J_2'^{\AFM}/J_2^{\AFM}$) for all the AA$'$VO(PO$_4)_2$ compounds along with the antiferromagnetic contributions to the exchange integrals $J_i^{\AFM}$ (in K) for AA$'$ = BaCd and SrZn
}
\begin{ruledtabular}
\begin{tabular}{cCCCCCC}
  AA$'$ & t_1 & t_1' & t_2 & t_2' & J_2'^{\AFM}/J_2^{\AFM} \\
  BaCd & 2 & -5 & 45 & 41 & 0.83 \\
  Pb$_2$ & 4\footnotemark & 15 & 47 & 38 & 0.67 \\
  BaZn & -11 & 12 & 46 & 36 & 0.61 \\
  SrZn & 4 & 17 & 43 & 26 & 0.37 \\\hline
   & J_1^{\AFM} & J_1'^{\AFM} & J_2^{\AFM} & J_2'^{\AFM} & J_1'^{\AFM}-J_1^{\AFM} \\
  BaCd & 0.05 & 0.3 & 21.0 & 17.4 & 0.25 \\ \footnotetext{The averaged NN coupling along the $b$ axis, see Sec.~\ref{structure} and Ref.~\onlinecite{foot4}}
  SrZn & 0.2 & 3.0 & 19.0 & 7.0 & 2.8 \\
\end{tabular}
\end{ruledtabular}
\end{table}

In Table~\ref{tab_hoppings}, we list the leading hopping parameters for all the AA$'$VO(PO$_4)_2$ compounds along with the resulting $J^{\AFM}$ values for the two representative and extreme cases, SrZnVO(PO$_4)_2$ and BaCdVO(PO$_4)_2$. All the hoppings beyond $t_2$ and $t_2'$ are negligible (below 2~meV). We find a notable distortion of the square lattice in all the materials under study, yet the magnitude of the distortion is rather different. The NNN couplings are AFM (see Table~\ref{tab_experiment}) and correspond to long V--V separations, hence the FM contributions are expected to be small, and one could consider the $J_2'^{\AFM}/J_2^{\AFM}$ ratio as a good estimate for the distortion ratio, $J_2'/J_2$, with respect to the ideal value $J_2'/J_2=1$ for the regular FSL. We find the least pronounced distortion ($J_2'^{\AFM}/J_2^{\AFM}=0.83$) in BaCdVO(PO$_4)_2$ and the strongest distortion ($J_2'^{\AFM}/J_2^{\AFM}=0.37$) in SrZnVO(PO$_4)_2$. 

The NN couplings are ferromagnetic (see Table~\ref{tab_experiment}), hence the TB model does not yield the direct estimate of $J_1'/J_1$. However, one can assume similar FM contributions to $J_1$ and $J_1'$ due to similar V--V separations. Then, the difference between the NN couplings should originate from the difference between $J_1^{\AFM}$ and $J_1'^{\AFM}$, and the value of $J_1'^{\AFM}-J_1^{\AFM}$ is the most convenient characteristic of the distortion. Both $J_1^{\AFM}$ and $J_1'^{\AFM}$ are well below 1 K in BaCdVO(PO$_4)_2$. In SrZnVO(PO$_4)_2$, $J_1'^{\AFM}$ is about 3 K, while $J_1^{\AFM}$ is still below 1 K. Thus, one can expect the sizable difference between $J_1$ and $J_1'$ in SrZnVO(PO$_4)_2$. In summary, we find weak distortion of the square lattice in BaCdVO(PO$_4)_2$ and a more pronounced distortion for AA$'$ = BaZn and Pb$_2$. The spin lattice of SrZnVO(PO$_4)_2$ is distorted with respect to both the NN and NNN couplings. Below, we will confirm this conclusion with the LSDA+$U$ calculations (see Table~\ref{tab_lsda+u}). 

\begin{figure*}
\includegraphics{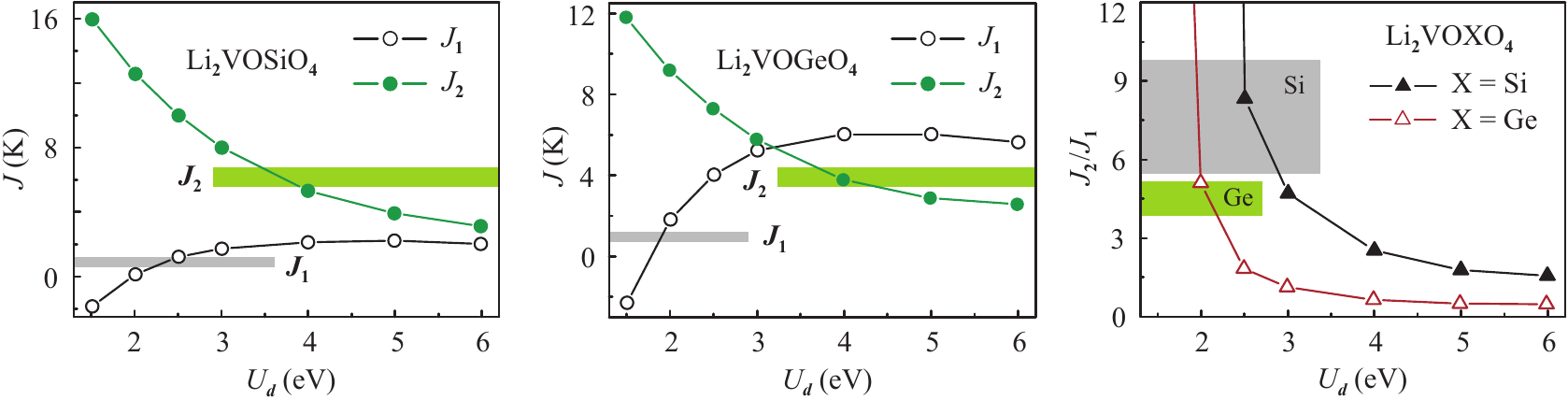}
\caption{\label{fig_u}(Color online)
Exchange couplings in Li$_2$VOSiO$_4$ (left panel) and Li$_2$VOGeO$_4$ (middle panel) calculated for different values of Coulomb repulsion parameter $U_d$ and the resulting frustration ratios $J_2/J_1$ for both the compounds (right panel). Shaded stripes show experimental estimates from Refs.~\onlinecite{helge2002,helge2003,misguich2003} and~\onlinecite{enrique}. 
}
\end{figure*}

The TB estimates are in reasonable agreement with the experimental data. We find the small NN hoppings consistent with the FM nature of $J_1^{\exp}$. On the other hand, the sizable hoppings $t_2$ and $t_2'$ imply AFM NNN couplings in agreement with the AFM coupling $J_2^{\exp}$. We further confirm the FM $J_1,J_1'$ -- AFM $J_2,J_2'$ scenario with the LSDA+$U$ calculations (see Table~\ref{tab_lsda+u}) that perfectly match the results of the TB analysis. The remarkable agreement of the TB and LSDA+$U$ results ensures sufficient accuracy of our approach. However, the calculated values of $J_2$ and $J_2'$ are likely overestimated as compared to $J_2^{\exp}$, the averaged NNN coupling (Table~\ref{tab_experiment}). We can speculate about several reasons for this discrepancy. 

First, one should not expect very precise results of band structure calculations while analyzing weak exchange couplings in AA$'$VO(PO$_4)_2$. The calculated $J$ values likely include a systematic error that, however, does not invalidate any of the qualitative conclusions: note the studies of the Li$_2$VOXO$_4$ (X = Si, Ge) compounds\cite{helge2002,helge2003} as an instructive example of the correct microscopic scenario, emerging from LDA calculations. Second, exchange couplings are highly sensitive to the geometry of superexchange pathways (see, e.g., Ref.~\onlinecite{ag2vop2o7}), hence it is essential to use the accurate structural information in order to obtain precise computational estimates. In case of AA$'$VO(PO$_4)_2$, the crystal structures are solved from single-crystal diffraction data, and the structural information should be precise. Yet, there are some unresolved issues (e.g., poorly reproducible superstructure reflections for Pb$_2$VO(PO$_4)_2$, see Ref.~\onlinecite{shpanchenko}) that can be crucial for the accuracy of the computational estimates. Along this paper, we mainly focus on the qualitative differences between the AA$'$VO(PO$_4)_2$ compounds, while the quantitative analysis should likely reference to the experimental data (see Sec.~\ref{discussion} for further discussion) and may require additional structural studies.
\begin{table*}
\renewcommand{\arraystretch}{1.5}
\caption{\label{tab_lsda+u}
LSDA+$U$ estimates of the exchange couplings in BaCdVO(PO$_4)_2$ and SrZnVO(PO$_4)_2$. $U_d$ is the Coulomb repulsion parameter of LSDA+$U$. The columns ($J_1'-J_1$) and $J_2'/J_2$ should be directly compared to the last column of Table~\ref{tab_hoppings}.
}
\begin{ruledtabular}
\begin{tabular}{cC@{\hspace{4ex}}CCC@{\hspace{4ex}}CCC}
  AA$'$ & U_d\text{ (eV)} & J_1\text{ (K)} & J_1'\text{ (K)} & J_1'-J_1\text{ (K)} & J_2\text{ (K)} & J_2'\text{ (K)} & J_2'/J_2 \\
  BaCd & & & & & & \\
  & 2.0 & -8.6 & -6.3 & 2.3 & 27 & 21.6 & 0.80 \\
  & 2.5 & -2.2 & -1.3 & 0.9 & 21.9 & 17.8 & 0.81 \\
  & 3.0 & 1.6 & 1.6 & 0.0 & 18.1 & 14.6 & 0.81 \\\hline
  SrZn & & & & & & \\
  & 2.0 & -10.0 & -5.4 & 4.6 & 24.8 & 11.3 & 0.46 \\
  & 2.5 & -2.4 & 1.6 & 4.0 & 20.8 & 8.7 & 0.42 \\
  & 3.0 & 2.7 & 6.3 & 3.6 & 17.5 & 7 & 0.40 \\
\end{tabular}
\end{ruledtabular}
\end{table*}

\subsection{LSDA+$U$ results}
To get a direct estimate of FM NN couplings in AA$'$VO(PO$_4)_2$, we turn to the LSDA+$U$ approach. In this approach, one has to select an appropriate value of $U_d$, the on-site Coulomb repulsion parameter (see Sec.~\ref{methods}).\cite{foot3} A number of previous works have established $U_d=3.5-4$~eV for several V$^{+4}$-containing compounds,\cite{korotin,boukhvalov-2003,boukhvalov-2004} although the higher value of $U_d=6$~eV\cite{V12-2006} as well as a lower value of $U_d=2.3$~eV\cite{na2v3o7} were proposed for specific materials. The reason for the higher $U_d$ value in Ref.~\onlinecite{V12-2006} was tentatively ascribed to the reduced $p-d$ hybridization. Still, there is one more uncertainty in the selection of this number. The constrained LDA procedure and the further LSDA+$U$ calculations are usually performed within the LMTO-ASA (linearized muffin-tin orbitals, atomic spheres approximation) approach that employs muffin-tin orbitals. In our work, we use a different basis set (atomic-like local orbitals of FPLO), hence the $U_d$ value may also be different, because the $U_d$ potential is applied to the orbitals from the basis set and depends on the particular choice of the $d$ functions. Indeed, our previous FPLO studies showed that the $U_d$ of 5 or even 6 eV was required to reproduce the magnetic interactions in a number of V$^{+4}$-containing phosphates.\cite{ag2vop2o7,cavpo} Keeping in mind the ambiguity of the $U_d$ choice due to the differences in the computational method (basis set) and in the structural features (different $p-d$ hybridization), we do not use any of the previously established $U_d$ values. Rather, we perform calculations for a broad range of $U_d$ values and use well-studied and structurally similar vanadium compounds as a reference.

According to Sec.~\ref{structure}, the structure of the magnetic layer in the AA$'$VO(PO$_4)_2$ phosphates is very similar to that of Li$_2$VOXO$_4$ compounds (X = Si, Ge). Exchange couplings in the latter materials are firmly established by fitting magnetic susceptibility and specific heat data with the HTSE.\cite{helge2002,helge2003,misguich2003,enrique} We calculate the exchange couplings in Li$_2$VOXO$_4$ for a wide range of $U_d$ values and compare the results with the experiment (Fig.~\ref{fig_u}). In the left and middle panels of Fig.~\ref{fig_u}, one can see that any reasonable value of $U_d$ yields the correct energy scale, but there is no unique $U_d$ value yielding accurate estimates of both $J_1$ and $J_2$. At low $U_d$, $J_2$ is overestimated, while high $U_d$ values tend to overestimate $J_1$. As long as we are interested in the frustration, the essential quantity is the frustration ratio, $J_2/J_1$. This quantity is found with a sizable error bar due to the uncertainty for the low $J_1$ values. Computational estimates of $J_2/J_1$ are also quite uncertain and highly sensitive to the $U_d$ value (see right panel of Fig.~\ref{fig_u}). At $U_d>3$ eV, we find very low frustration ratios, contradicting the $J_2\gg J_1$ regime established experimentally. The narrow range of $U_d=2-3$ eV is able to reproduce the reasonable frustration ratios, while the $U_d$ values below 2 eV lead to FM $J_1$. Thus, we argue that the $U_d$ value of $2-3$ eV should be optimal for reproducing exchange couplings in the layered vanadium FSL compounds. Yet one can construct the correct picture solely based on the LSDA+$U$ results only in a narrow range of the $U_d$ values.

In the LSDA+$U$ study, we restrict ourselves to BaCdVO(PO$_4)_2$ and SrZnVO(PO$_4)_2$ as the "edge" members of the AA$'$VO(PO$_4)_2$ series: these compounds show the least and the most pronounced distortion, respectively (see Table~\ref{tab_hoppings}). The LSDA+$U$ results are summarized in Table~\ref{tab_lsda+u}. The exchange couplings show a sizable dependence on $U_d$, even in the narrow range of $U_d=2-3$ eV. Nevertheless, we find the experimentally observed FM NN -- AFM NNN regime at $U_d=2.0$ and $2.5$ eV. Additionally, the numbers are in excellent agreement with the TB results, the difference between $J_2$ and $J_2'$ (quantified by the $J_2'/J_2$ ratio) as well as the different AFM contributions to $J_1$ and $J_1'$ (quantified by the difference $J_1'-J_1$, similar to Table~\ref{tab_hoppings}) are remarkably reproduced. 

The FSL-like spin system of the AA$'$VO(PO$_4)_2$ compounds includes four inequivalent exchange couplings. In case of BaCdVO(PO$_4)_2$, the respective NN ($J_1, J_1'$) and NNN ($J_2,J_2'$) couplings nearly match and give rise to the almost regular FSL. In BaZnVO(PO$_4)_2$ and Pb$_2$VO(PO$_4)_2$, the distortion is more pronounced. In case of SrZnVO(PO$_4)_2$, the square lattice is strongly distorted: the two NNN couplings differ by a factor of three, and the two NN couplings do not match as well. Using $J_1^{\exp}\simeq -8.3$~K as the averaged value (Table~\ref{tab_experiment}) and assuming $J_1'-J_1\simeq 3$ K (Table~\ref{tab_hoppings}), we estimate $J_1'/J_1\simeq 0.69$. In the next section, we discuss the structural origin of the spin lattice distortion.
 
\section{Structural origin of the distortion}
\label{factors}
To study the influence of individual structural factors on the spin lattice distortion in the AA$'$VO(PO$_4)_2$ compounds, we construct a number of model structures. The initial structure resembles that of $\alpha_{\text{I}}$-LiVOPO$_4$ (see Section~\ref{methods}) and includes regular vanadium and phosphorous polyhedra in the [VOPO$_4$] layers (the left panel of Fig.~\ref{fig_layer}). The cation--oxygen separations are 1.950 \r A for the basal oxygen atoms and 1.582 \r A for the axial oxygen atom in the VO$_5$ square pyramids and 1.543 \r A for the PO$_4$ tetrahedra. Then, we introduce certain distortions and analyze their influence on the magnetic interactions. We use the TB approach; therefore, we focus on the AFM NNN couplings and trace the change of the $J_2'/J_2$ ratio. In the end of the section, we briefly comment on the NN couplings and the $J_1'$ vs. $J_1$ distortion.

The difference between the AA$'$VO(PO$_4)_2$ compounds originates from different metal cations located between the [VOPO$_4$] layers. The cation size is reduced along the series from AA$'$ = BaCd to AA$'$ = SrZn, and this trend correlates to the reduction of $J_2'/J_2$ from $\simeq 0.8$ in BaCdVO(PO$_4)_2$ to $\simeq 0.4$ in SrZnVO(PO$_4)_2$. The coincidence of the two trends gives a hint that the cation size should be the origin of the distortion. Still, the influence of the cation size on the structure of the [VOPO$_4$] layers is quite complex. 

\begin{figure*}
\includegraphics[scale=0.9]{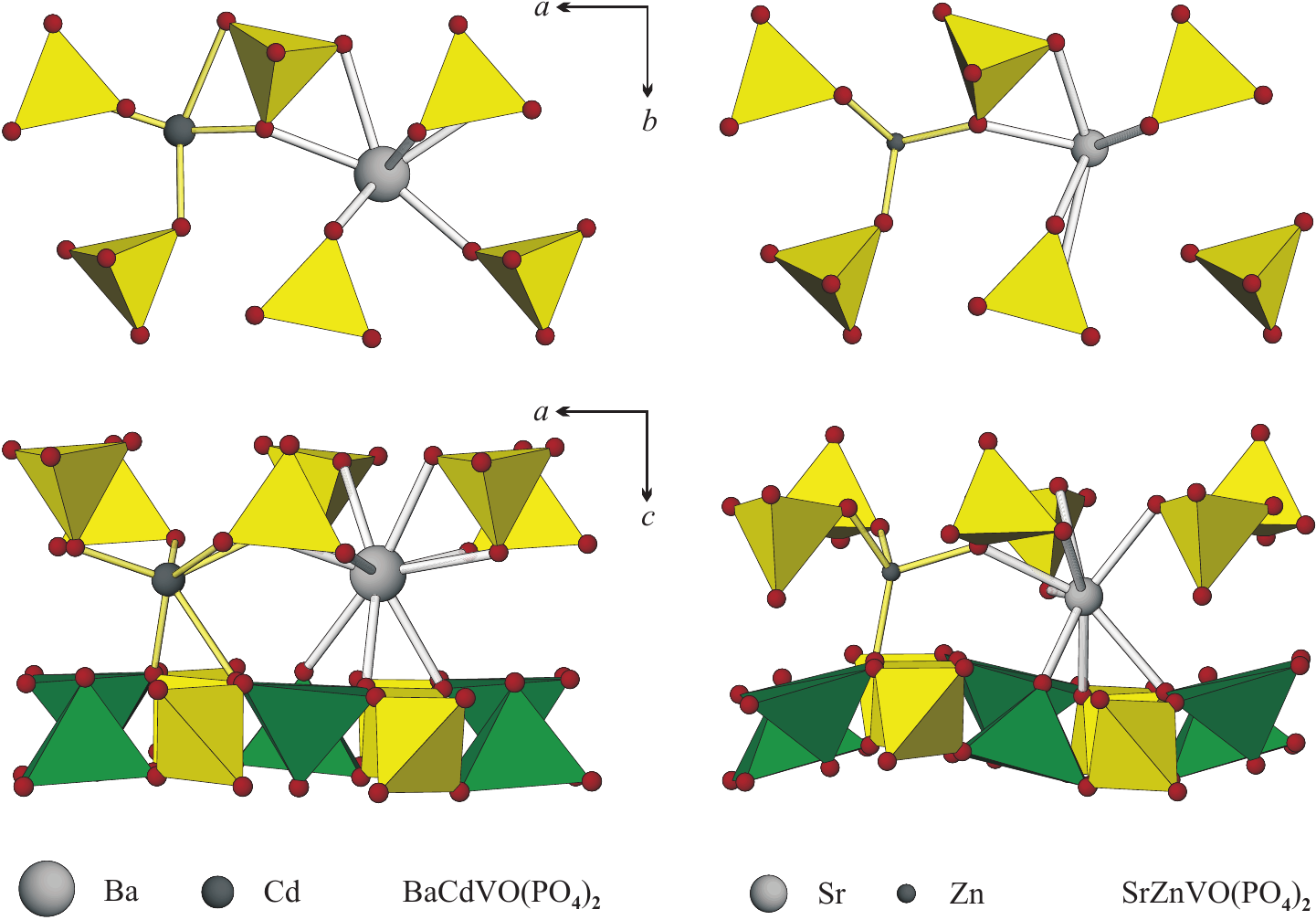}
\caption{\label{fig_interlayer}(Color online)
Crystal structures of BaCdVO(PO$_4)_2$ (left panels) and SrZnVO(PO$_4)_2$ (right panels): the upper panels show the interlayer [AA$'$PO$_4$] blocks, while the bottom panels present the buckling of the [VOPO$_4$] layers. The change of Ba and Cd for Sr and Zn leads to the reduction of the coordination numbers and the resulting reorganization of the structure: the layer buckling and the stretching in the $ab$ plane (see text for details).
}
\end{figure*}

Smaller metal cations tend to have shorter metal--oxygen distances and lower coordination numbers. Thus, barium is surrounded by nine oxygen atoms in BaZnVO(PO$_4)_2$ and BaCdVO(PO$_4)_2$, while strontium has only eight neighboring oxygens in SrZnVO(PO$_4)_2$. The same holds for cadmium (coordination number of~6) vs. zinc (coordination number of~4). Larger Ba and Cd cations are compatible with the nearly flat [VOPO$_4$] layers (see the left panel of Fig.~\ref{fig_interlayer}). To provide proper oxygen coordination for smaller cations, the layers have to buckle. The Sr and Zn cations occupy the positions at the points of the downward and upward curvature, respectively (see the right bottom panel of Fig.~\ref{fig_interlayer}). The layer buckling can be quantified via the angle $\varphi$, see Table~\ref{tab_geometry} and the middle panel of Fig.~\ref{fig_layer}. 

The structural changes are not confined to the layer buckling. In particular, the unit cell parameters change in a peculiar manner. As the cation size is decreased, the $a$ and $b$ parameters (in-layer spacings) are increased (see Table~\ref{tab_geometry}), while the $c$ parameter (interlayer spacing) is decreased to provide the overall reduction of the unit cell volume expected for the substitution by a smaller cation. The increase of the in-layer dimensions can also be understood via the change of the cation coordination numbers. Barium cations coordinate oxygen atoms from four surrounding PO$_4$ tetrahedra in the [AA$'$PO$_4$] interlayer block (upper left panel of Fig.~\ref{fig_interlayer}). Strontium cations coordinate three tetrahedra only (upper left panel of Fig.~\ref{fig_interlayer}), while the fourth tetrahedron is "pushed away", thus expanding the unit cell along the $a$ and $b$ directions. 

\begin{figure}
\includegraphics{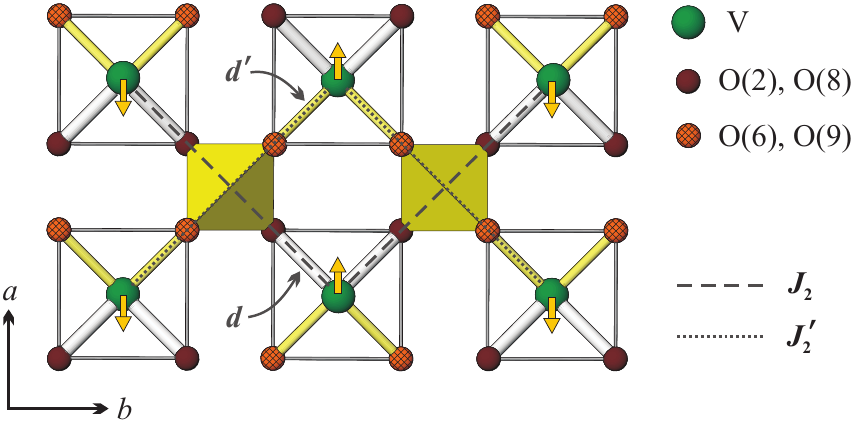}
\caption{\label{fig_displacement}(Color online)
Distortion of the [VOPO$_4$] layer as implemented in the model structures. Arrows show displacements of the vanadium atoms away from their ideal positions in the centers of the VO$_5$ square pyramids. The displacements yield two different V--O distances ($d$ and $d'$) and two inequivalent NNN couplings ($J_2$ and $J_2'$). Small spheres with solid and hatched filling denote different types of oxygen atoms with shorter (white bonds) and longer (shaded bonds) V--O distances, respectively. Dashed and dotted lines indicate the interactions $J_2$ and $J_2'$.
}
\end{figure}

The picture presented in the last two paragraphs and visualized in Fig.~\ref{fig_interlayer} does not reflect all the structural changes including, e.g., slight shifts of the metal cations and tiltings of the PO$_4$ tetrahedra. However, this picture grasps the essential changes that bear influence on the magnetic [VOPO$_4$] layers and on the exchange couplings. There are two main structural changes in the magnetic layers: (i) the buckling and (ii) the stretching in the $ab$ plane. The latter is performed via the distortion of the VO$_5$ pyramids, while the PO$_4$ tetrahedra remain rigid. The specific distortion of the square pyramids is shown in Fig.~\ref{fig_displacement}: vanadium atoms are shifted away from the pyramid center and yield two longer and two shorter V--O bonds. Two types of the V--O bonds are easily distinguished in Table~\ref{tab_geometry}. The V--O(2) and V--O(8) distances remain nearly constant ($1.95-2.00$ \r A) in the whole AA$'$VO(PO$_4)_2$ series, while the two other distances [\mbox{V--O(6)} and \mbox{V--O(9)}] are expanded from about 2.00 \r A in BaCdVO(PO$_4)_2$ up to 2.11 \r A in SrZnVO(PO$_4)_2$. Both the buckling and the stretching of the magnetic layers are most pronounced in SrZnVO(PO$_4)_2$. Below, we construct proper model structures and separately analyze the influence of these effects on the exchange couplings.

\begin{table*}
\begin{minipage}{\textwidth}
\renewcommand{\arraystretch}{1.5}
\caption{\label{tab_geometry}
Lattice parameters ($a,b,c$) and some other geometrical characteristics of the AA$'$VO(PO$_4)_2$ compounds. The V--O distances are given in units of \r A, and the oxygen positions are numbered according to the structural data in Refs.~\onlinecite{srznvp2o9} and \onlinecite{baznvp2o9}. The angle $\varphi$ is a measure for the buckling of the [VOPO$_4$] layers as shown in Fig.~\ref{fig_layer}.
}
\begin{ruledtabular}
\begin{tabular}{cCCCCCCCCc}
  AA$'$ & a\text{ (\r A)} & b\text{ (\r A)} & c\text{ (\r A)} & \varphi (^{\circ}) & \text{V--O(2)} & \text{V--O(8)} & \text{V--O(6)} & \text{V--O(9)} & Ref. \\
  BaCd & 8.838 & 8.915 & 19.374 & 172 & 1.977 & 1.975 & 2.011 & 1.992 & \onlinecite{srznvp2o9} \\
  Pb$_2$\footnote{Non-standard setting used, see Sec.~\ref{structure}} & 9.016 & 8.747 & 9.863\footnotemark & 155 & 1.954 & 1.975 & 2.024 & 2.000 & \onlinecite{shpanchenko} \footnotetext{The unit cell of Pb$_2$VO(PO$_4)_2$ includes one magnetic layer in contrast to the other AA$'$VO(PO$_4)_2$ compounds with two [VOPO$_4$] layers in the unit cell. Yet the $c$ parameter is not the true (shortest) interlayer distance due to the monoclinic symmetry of the structure.}\\ 
  BaZn & 8.814 & 9.039 & 18.538 & 160 & 1.956 & 1.974 & 2.045 & 1.993 & \onlinecite{baznvp2o9} \\
  SrZn & 9.066 & 9.012 & 17.513 & 150 & 1.971 & 1.999 & 2.110 & 2.039 & \onlinecite{srznvp2o9} \\
\end{tabular}
\end{ruledtabular}
\end{minipage}
\end{table*}

To reproduce the layer buckling (i), we keep the VO$_5$ and PO$_4$ polyhedra rigid and simply change the \mbox{V--O--P} angles to achieve the necessary buckling angle $\varphi$. To reproduce the layer stretching (ii), vanadium atoms are shifted away from the centers of the pyramids (see Fig.~\ref{fig_displacement}), and the unit cell is properly expanded in the $ab$ plane. Then, two V--O distances remain constant ($d=1.95$ \r A), while two other distances ($d'$) are increased. The expansion in quantified by the value of $\Delta d=d'-d$. 

\begin{figure}
\includegraphics[scale=0.9]{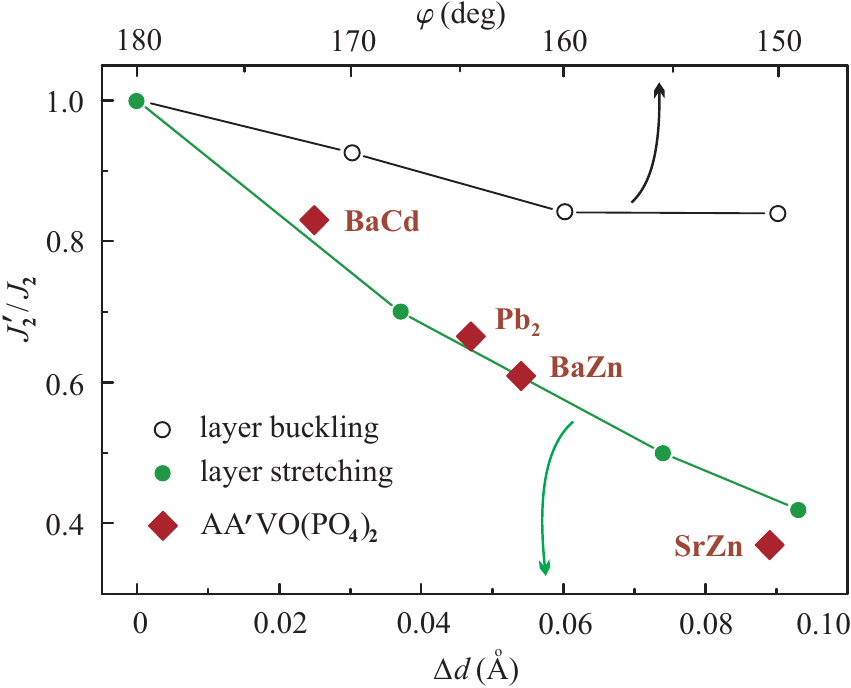}
\caption{\label{fig_distortion}(Color online)
Spin lattice distortion ($J_2'/J_2$) for the model structures with the layer buckling (empty circles) and the layer stretching (filled circles) and for the real AA$'$VO(PO$_4)_2$ compounds (filled diamonds). The layer buckling is quantified by the buckling angle $\varphi$ (Fig.~\ref{fig_layer}). The layer stretching is imposed by shifting vanadium atoms and quantified by $\Delta d=d'-d$ (see Fig.~\ref{fig_displacement}). For the real compounds, $d=(d_{\text{V--O(2)}}+d_{\text{V--O(8)}})/2$ and $d'=(d_{\text{V--O(6)}}+d_{\text{V--O(9)}})/2$ (see Table~\ref{tab_geometry}).
}
\end{figure}
\begin{figure*}
\includegraphics{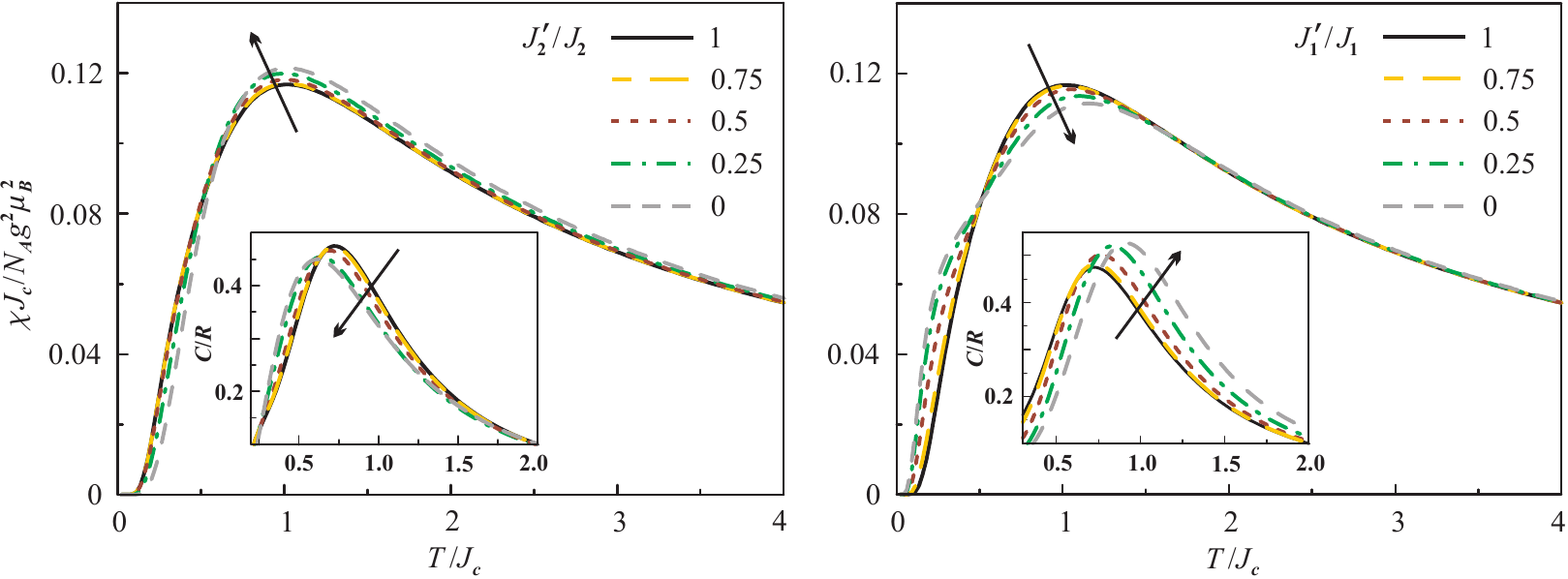}
\caption{\label{fig_simulation}(Color online)
Full diagonalization results for the distorted FSL model: magnetic susceptibility (primary figures) and specific heat (insets). The simulations are performed at fixed averaged couplings $\bar J_1=(J_1+J_1')/2$ and $\bar J_2=(J_2+J_2')/2$ and the fixed frustration ratio $\bar J_2/\bar J_1=-2$. In the left and right panels, $J_2$ and $J_2'$ or $J_1$ and $J_1'$ are varied, respectively. The arrows show the changes upon increasing the distortion. The thermodynamic energy scale $J_c$ is defined as $J_c=\sqrt{(J_1+J_1'^2+J_2+J_2'^2)/2}$.
}
\end{figure*}

In Fig.~\ref{fig_distortion}, we plot the distortion of the NNN couplings ($J_2'/J_2$) as found for the real compounds and for the model structures. We find that the buckling of the layers is unable to account for the spin lattice distortion observed in most of the AA$'$VO(PO$_4)_2$ compounds (see open circles in Fig.~\ref{fig_distortion}). On the other hand, the shifts of the vanadium atoms and the resulting layer stretching perfectly reproduce the distortion, even without considering the layer buckling (see filled circles in Fig.~\ref{fig_distortion}). Thus, we conclude that the distortion of the NNN couplings in AA$'$VO(PO$_4)_2$ originates from the distortion of the VO$_5$ pyramids. We can also unambiguously assign the weaker interaction $J_2'$ to the longer V--O(6) and V--O(9) separations consistent with the LSDA+$U$ results for the real compounds.

Unfortunately, the trends for the NN couplings are less clear. According to the discussion in Sec.~\ref{distortion}, the difference between $J_1$ and $J_1'$ originates from different AFM contributions to these couplings. The initial model structure does not show any sizable NN hoppings (both $t_1$ and $t_1'$ are below 4 meV), and the shifts of the vanadium atoms within the flat layer do not change these hoppings. The layer buckling enlarges $t_1$ and $t_1'$ up to $10-15$ meV, i.e., the TB results for BaZnVO(PO$_4)_2$ are reproduced (see Table~\ref{tab_hoppings}). However, the model structures do not show the anisotropy of the NN couplings, as found in SrZnVO(PO$_4)_2$ and Pb$_2$VO(PO$_4)_2$. The anisotropy is likely caused by more subtle changes that are not included in our model structures. Nevertheless, the flat [VOPO$_4$] layer [as found in BaCdVO(PO$_4)_2$] does not show any considerable AFM contributions to the NN couplings, hence the anisotropy of the NN couplings in the flat-layer compounds should be small.

\section{Extended FSL model}
\label{simulation}
In this section, we discuss the properties of the extended FSL model. We address thermodynamic properties, magnetic susceptibility and specific heat, because these quantities are measured experimentally and commonly used for the evaluation of the exchange couplings in the FSL compounds. The extended model includes four independent parameters, but we are mainly interested in the role of the distortion, i.e., the difference between $J_1$ and $J_1'$ or $J_2$ and $J_2'$. Therefore, we fix the averaged NN and NNN couplings ($\bar J_1$ and $\bar J_2$, respectively) and the effective frustration ratio $\alpha=\bar J_2/\bar J_1$. Then we vary either $J_1$ and $J_1'$ or $J_2$ and $J_2'$. We performed the simulations for two representative values, $\alpha=-2$ and $\alpha=-1$, to study the frustration regime relevant for Pb$_2$VO(PO$_4)_2$ and BaZnVO(PO$_4)_2$ or SrZnVO(PO$_4)_2$ and BaCdVO(PO$_4)_2$, respectively (see Table~\ref{tab_experiment}). In the following, we present the results obtained at $\alpha=-2$ (Fig.~\ref{fig_simulation}). The simulations for $\alpha=-1$ reveal a very similar behavior, thus we do not discuss them in detail. 

The results for the regular FSL at $\alpha=-2$ match that of Ref.~\onlinecite{shannon2004} (for the comparison, one should use the frustration angle $\varphi_f/\pi\simeq 0.65$ with $\varphi_f=\text{tan}^{-1}\alpha$). The distortions of the NN and NNN bonds have different effects on the thermodynamic properties. The distortion of the NNN bonds ($J_2'$ vs. $J_2$, left panel of Fig.~\ref{fig_simulation}) leads to a slight shift of the susceptibility maximum to lower temperatures, while the absolute value at the maximum is increased. The maximum of the specific heat is also shifted to lower temperatures, but its height is reduced. The distortion of the NN bonds ($J_1'$ vs. $J_1$, right panel of Fig.~\ref{fig_simulation}) leads to opposite and more pronounced changes. The maxima are shifted to higher temperatures, the susceptibility maximum is reduced, while the specific heat maximum is increased. 

To understand these results, one should recall that the magnetic susceptibility and the specific heat maxima are caused by correlated spin excitations. These excitations compete with quantum fluctuations caused by the low dimensionality and the magnetic frustration. Thus, the shift of the maxima to lower/higher temperatures implies the enhancement/reduction of the quantum fluctuations. Then the effects for the susceptibility and the specific heat are opposite, since spin correlations contribute to the specific heat and increase its value, but lead to AFM ordering, hence reducing the susceptibility. The trends presented in Fig.~\ref{fig_simulation} suggest that the distortion of the NNN bonds enhances quantum fluctuations (see the explanation below), while the distortion of the NN bonds reduces the fluctuations, and the latter effect is more pronounced.

The effect of the NN bonds distortion is consistent with the previous theoretical results for the spatially anisotropic FSL model.\cite{starykh2004,sindzingre2004,bishop2008} The narrowing and closing of the critical (spin-liquid) region corresponds to the reduction of the quantum fluctuations, as observed in the thermodynamic data. One can get further insight into this effect by considering energies of the ordered structures within the classical Heisenberg model. At $\alpha=-2$, the competing ground states are the FM and columnar AFM ordering (see Fig.~\ref{fig_diagram}). The columnar AFM state is favored by AFM NNN interactions and by the FM interaction along the directions of columns (say, along the $b$ axis, i.e., the respective interaction is $J_1$). Yet, the FM interaction $J_1'$ along the $a$ axis is unfavorable for the columnar ordering. As the absolute value of $J_1'$ is reduced and that of $J_1$ is increased, the columnar AFM state is stabilized, and frustration is released. Applying similar considerations to the NNN bonds distortion, we find that the distortion does not change the energies of the FM and columnar AFM states, hence the magnitude of the frustration should remain unchanged. This conclusion is consistent with the relatively weak effect of the NNN bonds distortion. Still, the certain enhancement of the frustration is clearly visible in the thermodynamic data and likely related to quantum effects. We can speculate that the reduction of the $J_2'/J_2$ ratio leads to the formation of spin chains within the 2D lattice. The chains are formed by the $J_2$ bonds and run along the $b$ axis (see the right panel of Fig.~\ref{fig_layer}). Then, this one-dimensional feature of the spin system should enhance quantum fluctuations due to the effectively reduced dimensionality.

Now, we turn to the case of moderate distortion relevant for the AA$'$VO(PO$_4)_2$ compounds. According to Sec.~\ref{distortion}, the strongest distortion is found in SrZnVO(PO$_4)_2$ with $J_1'/J_1\simeq 0.7$ and $J_2'/J_2\simeq 0.4$. The respective susceptibility curves in Fig.~\ref{fig_simulation} nearly match that for the regular FSL. Thus, the fitting of the experimental susceptibility data should yield averaged exchange couplings of the distorted FSL, $\bar J_1$ and $\bar J_2$. This conclusion provides a reliable basis for the interpretation of the experimental values listed in Table~\ref{tab_experiment}. The changes in the specific heat curves are also minor, hence the experimental specific heat will be described by the HTSE. Thus, the thermodynamic properties of the AA$'$VO(PO$_4)_2$ compounds should fit the regular FSL model with averaged exchange parameters as an excellent approximation, and this is the case.\cite{kaul2004,enrique,bacdvp2o9}

\section{Discussion and summary}
\label{discussion}
In this study, we performed a detailed microscopic investigation of the distorted FSL spin systems in the AA$'$VO(PO$_4)_2$ compounds. We estimated the magnitude of the distortion, found the structural origin of the distortion and analyzed the thermodynamic properties of the extended FSL model. Below, we consider the consequences of these findings in several aspects: (i) the physics of the AA$'$VO(PO$_4)_2$ phosphates; (ii) layered vanadium compounds as a playground for the search of new FSL materials; and (iii) lattice distortion in frustrated spin systems.

The basic experimental results for the AA$'$VO(PO$_4)_2$ compounds can be interpreted within the framework of the regular FSL model.\cite{kaul2004,enrique,bacdvp2o9} According to Sec.~\ref{simulation}, thermodynamic measurements lead to relevant, averaged exchange couplings $J_1^{\exp}$ and $J_2^{\exp}$. These values place all the AA$'$VO(PO$_4)_2$ compounds to the columnar AFM region of the FSL phase diagram (Fig.~\ref{fig_diagram}). Indeed, neutron scattering results for Pb$_2$VO(PO$_4)_2$ and SrZnVO(PO$_4)_2$ (Refs.~\onlinecite{skoulatos2007} and~\onlinecite{skoulatos2008}) confirm the columnar ordering. Yet, the experimental situation is not fully clear, since nuclear magnetic resonance (NMR) and muon spin relaxation ($\mu$SR) studies suggest a broad distribution of local magnetic fields in the ordered phase, hence pointing to a more complex ground state (e.g., incommensurate ordering), at least for some of the AA$'$VO(PO$_4)_2$ compounds.\cite{ramesh} One can suggest that the ground state of these materials is influenced by the spin lattice distortion. Although detailed investigation of the ground state of the extended FSL model lies beyond the scope of the present study, the introduction to the extended model and the evaluation of the model parameters is a first step toward understanding the ground state of AA$'$VO(PO$_4)_2$. Further theoretical and experimental (neutron scattering, NMR, $\mu$SR) studies on the ground state properties are highly desirable and should be stimulated by our work. 

Apart from the thorough studies of the ground state, one can suggest a more simple way for the experimental observation of the spin lattice distortion in AA$'$VO(PO$_4)_2$. According to Sec.~\ref{simulation}, the distortion of the nearest-neighbor bonds ($J_1$ vs. $J_1'$) stabilizes the columnar AFM state with respect to the FM state. Within the classical model, the energy difference between the FM and columnar AFM states corresponds to the saturation field. Therefore, the $J_1\neq J_1'$ scenario should have an effect on the saturation field. In case of BaCdVO(PO$_4)_2$ with the nearly regular FSL, the saturation field is in excellent agreement with the averaged couplings $J_1^{\exp}$ and $J_2^{\exp}$ (Ref.~\onlinecite{bacdvp2o9}). However, the saturation field of SrZnVO(PO$_4)_2$ should be different from the field estimated using $J_1^{\exp}$ and $J_2^{\exp}$. High-field magnetization studies of the AA$'$VO(PO$_4)_2$ compounds will challenge this proposition and enable the quantitative analysis of the spin lattice distortion in AA$'$VO(PO$_4)_2$.

The next important issue is the capability of layered vanadium compounds to reveal new strongly frustrated FSL materials. The basic structural element of these compounds is the [VOXO$_4$] layer shown in the left panel of Fig.~\ref{fig_layer}. In this layer, non-magnetic XO$_4$ tetrahedra mediate either FM or weak AFM NN and AFM NNN interactions. Then there are two ways to reach the strongly frustrated regime of $\alpha\simeq -0.5$: one should either increase the absolute value of $J_1$ [as observed in SrZnVO(PO$_4)_2$] or decrease $J_2$ [as observed in BaCdVO(PO$_4)_2$]. According to the results of our study (Sec.~\ref{factors}), the NNN interactions ($J_2$) are sensitive to the V--O distances, since the magnitude of any superexchange interaction depends on the orbital overlap. Thus, it is possible to reduce the $J_2$ value by increasing the V--O separations in the VO$_5$ square pyramids. The factors influencing on the value of $J_1$ are less clear. In the AA$'$VO(PO$_4)_2$ compounds, the $J_1$ values correlate with the distortion of the NNN bonds. One may suggest the layer buckling or the distortion of the VO$_5$ pyramids as possible reasons for the increase of $J_1^{\exp}$ from BaCdVO(PO$_4)_2$ to SrZnVO(PO$_4)_2$. However, this conclusion is rather empirical, and other structural factors may be relevant as well.

The proper FSL material should combine the strong frustration with the lack of the distortion or, at least, with a relatively weak distortion of the spin lattice. It is also desirable to find a family of isostructural FSL compounds. Then, the replacement of the metal cations may facilitate the tuning of the system toward the strongly frustrated regime. Our results provide a clear recipe for the search of undistorted FSL materials. To get a regular FSL, one should keep the magnetic layer flat and avoid the distortion of the VO$_5$ square pyramids. Clearly, it is quite difficult to fulfill these criteria within the AA$'$VO(PO$_4)_2$ series. Different metal cations require different coordination numbers (see Fig.~\ref{fig_interlayer}); therefore, most of the respective compounds reveal a distorted FSL. In BaCdVO(PO$_4)_2$, the cation sizes are optimal to yield nearly flat and regular magnetic layers and to result in the weak distortion of the FSL. Though tuning is possible (see Table~\ref{tab_experiment}), most of the metal cations lead to a layer distortion and to a spin lattice distortion as well. 

To avoid the spin lattice distortion within a compound family, one can try to reduce the number of metal cations and to look for another filler of the interlayer space. For example, one can consider the A(VOPO$_4)_2\cdot 4$H$_2$O compounds (A = Ca, Cd, Sr, Pb, Ba, and Mg/Zn)\cite{lii1992,kang1992,papoutsakis1996,roca1997,roca1998,lefur1999,lefur1999-2,pbvpo} that reveal layered structures with [VOPO$_4$] layers separated by metal cations and water molecules. The resulting symmetry is orthorhombic, monoclinic, or even triclinic, but the respective spin lattice distortion should be weak. The layers are nearly flat for all the six compounds reported, and the distortion of the VO$_5$ pyramids is also negligible. Yet, the change of the metal cation bears influence on the magnetic interactions, as indicated by different Curie-Weiss temperatures.\cite{papoutsakis1996,roca1998,lefur1999,lefur1999-2,pbvpo} The magnitude of the frustration in the A(VOPO$_4)_2\cdot 4$H$_2$O compounds remains unknown. All the reports available consider NN interactions only. Clearly, this scenario is oversimplified, and one has to apply the FSL (rather than a simple square lattice) model while analyzing magnetic properties of A(VOPO$_4)_2\cdot 4$H$_2$O. Further studies of these materials should be very promising.

The last, but not least, point deals with the influence of the distortion on the properties of frustrated spin systems. In this study, we exemplified this problem by considering the distortion of the FSL. We found that different types of distortion had different effects on the magnitude of the frustration. The distortion of the NN bonds stabilizes the columnar AFM ordering and releases the frustration. Yet, the distortion of the NNN bonds keeps the strong frustration and even enhances the quantum fluctuations. These results suggest that both regular and distorted frustrated materials should be considered as proper realizations of the frustrated spin models. The case of the AA$'$VO(PO$_4)_2$ compounds shows how the distorted FSL materials can be successfully treated within the regular FSL model. In these compounds, the thermodynamic properties at available temperatures nearly match those of the regular model. However, the ground state properties may be different. We believe that the study of frustrated materials with spin lattice distortion will improve our understanding of the frustrated spin systems and facilitate the observation of interesting phenomena suggested by theory. 

In conclusion, we have studied the distorted frustrated square lattice in layered vanadium phosphates AA$'$VO(PO$_4)_2$ (AA$'$ = Pb$_2$, SrZn, BaZn, and BaCd). In these compounds, both nearest-neighbor and next-nearest-neighbor bonds of the square lattice are distorted, hence an extended spin model with four inequivalent exchange couplings should be considered. Our estimates of the individual model parameters suggest the least pronounced distortion in BaCdVO(PO$_4)_2$ and the most pronounced distortion in SrZnVO(PO$_4)_2$. The difference between the nearest-neighbor and next-nearest-neighbor interactions in these compounds originates from peculiar structural changes upon substituting interlayer metal cations A and A$'$. The buckling of the magnetic layers may change the interactions along the side of the square, while the distortion of the vanadium polyhedra and the resulting stretching of the magnetic layer lead to the difference between the diagonal interactions of the distorted square lattice. The distortion of the square lattice in AA$'$VO(PO$_4)_2$ is moderate from the point of view of the thermodynamic properties. The temperature dependences of the magnetic susceptibility and the specific heat resemble those of the regular square lattice, hence previous experimental studies of AA$'$VO(PO$_4)_2$ reported averaged couplings. These couplings can be used to place the compounds on the phase diagram of the regular model. In contrast to the moderate influence on the thermodynamic properties, the spin lattice distortion may have a larger effect on the ground state. Further experimental and theoretical studies of this problem are highly desirable.

\begin{acknowledgments}
The authors are grateful to Christoph Geibel for fruitful discussion. Financial support of GIF (I-811-257.14/03), RFBR (07-03-00890), and the Emmy Noether Program of the DFG is acknowledged. A.Ts. is grateful to MPI CPfS for hospitality and financial support during the stay.
\end{acknowledgments}


\begin{thebibliography}{64}
\bibitem[{foo({\natexlab{b}})}]{foot7}
\bibinfo{note}{Note that the exchange constants of the order of 10~K are at
  least 9 orders of magnitude smaller than the total energy per unit cell.}
\bibitem[{foo({\natexlab{d}})}]{foot4}
\bibinfo{note}{Assuming similar FM contributions to all the NN couplings, one
  can suggest that the very small $t_1$ in Pb$_2$VO(PO$_4)_2$ implies similar
  (and FM) interactions along the $b$ axis, despite two different superexchange
  pathways are present (see Sec.~\ref{structure}). This conclusion can be
  further supported by LSDA+$U$ calculations (not shown) and justifies our
  consideration of Pb$_2$VO(PO$_4)_2$ as the close analog of the other
  AA$'$VO(PO$_4)_2$ compounds.}
\bibitem[{foo({\natexlab{e}})}]{foot3}
\bibinfo{note}{Note that $U_d$, the Coulomb repulsion parameter of LSDA+$U$, is
  applied to atomic orbitals. This is different from $U_{\eff}$, the effective
  on-site Coulomb repulsion in the mixed vanadium-oxygen band. In general, the
  $U_d$ and $U_{\eff}$ values do not coincide.}
\end{thebibliography}
\end{document}